\documentclass[a4paper]{article}
\usepackage{amsmath,amssymb}
\title{The Scalar-flat K\"ahler Metric and Painlev\'e III}
\author{Shoji Okumura (okumura@math.sci.osaka-u.ac.jp)\\Math. Sci. Osaka Univ., Japan}
\date{July 20, 2000}

\begin{document}
\maketitle

\newtheorem{thm}{Theorem}[section]
\newtheorem{prop}[thm]{Proposition}
\newtheorem{rem}[thm]{Remark}
\newtheorem{definition}[thm]{Definition}
\newtheorem{cor}[thm]{Corollary}
\newtheorem{lemma}[thm]{Lemma}

\begin{abstract}

We study the anti-self-dual equation for non-diagonal $SU(2)$-invariant metrics and 
give an equivalent ninth-order system. This system reduce to a sixth-order system
if the metric is in the conformal class of scalar-flat-K\"ahler metric. 

\end{abstract}

\section{introduction}

The aim of this paper is to analyse scalar-flat K\"ahler metrics $g$ in real dimmension four
admitting an isometric action of $SU(2)$ with generically three-dimensional orbits.
A scalar-flat K\"ahler metric is a metric with zero scalar curvature which is K\"ahlerian with
respect to a complex structure on $M$. 
It is automatically anti-self-dual with respect to the canonical orientation.

Hitchin\cite{Hitchin} shows that
the $SU(2)$-invariant anti-self-dual metric is generically specified 
by a solution
of Painlev\'e VI type equation, and if the metric is scalar-flat K\"ahler
it is specified by a solution of Painlev\'e III type equation.
Hitchin used the twistor correspondence to associate the anti-self-dual
equation and Painlev\'e equation.  The lifted action of $SU(2)$
determines a pre-homogenious action of $SU(2)$ on the twistor
space $Z$, and it determines a isomonodromic family of connections
on $\mathbb{CP}^1$, and then we have Painlev\'e equations.
In this way, Dancer \cite{Dancer} analyse the scalar-flat K\"ahler metric with
$SU(2)$-symmetry.  In \cite{Hitchin}, Hichin obtains a complete classification
of anti-self-dual Einstein metrics admitting an isometric action of $SU(2)$ with 
three-dimensional orbits.  For the completeness analysis, it
is impotant to have the explicit form of anti-self-dual equation.

If the metric is diagonal, the explicit form of anti-self-dual
 equation is known, but if the metric is non-diagonal, it is
 known very little.  For scalar-flat K\"ahler metric, complex
structure is not known for non-diagonal metric.
 
In Section \ref{ODE} we show how non-diagonal metric is represented,
and by use of the block form of curvature tensor given by Besse
\cite{Besse}, we have the ninth-order system equivalent to 
the anti-self-dual equation.

In Section \ref{Painleve} we establishes the relationship between
$SU(2)$-invariant anti-self-dual manifold and the isomonodromic
deformation.  It is essentially equivalent to Hitchin's ansatz.
Still in our way, we have the explicit form of the isomonodromic
deformations, and we have the condition that the corresponding
Painlev\'e equation is type III.

In Section \ref{Hermitian}, we show that the anti-self-dual 
equation reduce to Painlev\'e III if and only if the metric
admits an Hermitian structure.  In this case, the anti-self-dual
equation is equivalent to a seventh-order system, and it also
admits K\"ahler structure, the seventh-order system reduce to
a sixth-order system.

\textbf{Acknowledgements}\\
The author would express his sincere gratitude to Professor Yousuke Ohyama, who
intoroduced him to the subject, for enlightening discussions.

\section{The non-diagonal anti-self-dual equations} \label{ODE}

We can write the $SU(2)$-invariant metric in the form
\begin{align*}
g=f(\tau) d\tau^2 + \sum _{l,m=1}^3 h_{l\,m}(\tau)\, \sigma_l \sigma_m,
\end{align*}
where $\{\sigma_1,\sigma_2,\sigma_3\}$ is a basis of left invariant one-forms on each
$SU(2)$-orbit satisfying
\begin{align*}
d\sigma_1&=\sigma_2\wedge\sigma_3, 
      & d\sigma_2 &=\sigma_3\wedge\sigma_1, & d\sigma_3=\sigma_1\wedge \sigma_2.
\end{align*}
Using the Killing form, we can diagonalize the metric $g$ on each $SU(2)$-orbit. 
Then we can express the metric as follows:
\begin{align*}
g=(a b c)^2 dt^2+a^2d\tilde{\sigma}_1^2+b^2\tilde{\sigma}_2^2+c^2\tilde{\sigma}_3^2,
\end{align*}
where $t=t(\tau), a=a(t), b=b(t), c=c(t)$ and 
\begin{align*}
\left(
  \begin{array}{c}
    \tilde{\sigma}_1   \\
    \tilde{\sigma}_2   \\
    \tilde{\sigma}_3   
  \end{array}
\right)=R(t)
\left(
  \begin{array}{c}
    \sigma_1   \\
    \sigma_2   \\
    \sigma_3   
  \end{array}
\right),
\end{align*}
$R(t)$ is $SO(3)$-valued function.

Since $\dot{R}R^{-1}$ (where $\dot{*}=\frac{d}{dt}$) is $\mathfrak{so}(3)$-valued, 
we can write
\begin{align*}
d\left(
  \begin{array}{c}
    \tilde{\sigma}_1   \\
    \tilde{\sigma}_2   \\
    \tilde{\sigma}_3   
  \end{array}
\right)&=R(t)
\left(
  \begin{array}{c}
    \sigma_2\wedge\sigma_3   \\
    \sigma_3\wedge\sigma_1   \\
    \sigma_2\wedge\sigma_2   
  \end{array}
\right)+
\dot{R}\, dt\wedge\left(
  \begin{array}{c}
    \sigma_1   \\
    \sigma_2   \\
    \sigma_3   \\
  \end{array}
\right)\\
 &=\left(
  \begin{array}{c}
    \tilde{\sigma}_2\wedge\tilde{\sigma}_3   \\
    \tilde{\sigma}_3\wedge\tilde{\sigma}_1   \\
    \tilde{\sigma}_1\wedge\tilde{\sigma}_2   \\
  \end{array}
\right)+
\left(
  \begin{array}{rrr}
    0   &  \xi_3  & -\xi_2   \\
    -\xi_3   & 0   &   \xi_1 \\
    \xi_2   & -\xi_1   & 0   \\
  \end{array}
\right)dt\wedge
\left(
  \begin{array}{c}
    \tilde{\sigma}_1   \\
    \tilde{\sigma}_2   \\
    \tilde{\sigma}_3   \\
  \end{array}
\right),
\end{align*}
for some $\xi_1=\xi_1(t)$, $\xi_2=\xi_2(t)$, $\xi_3=\xi_3(t)$.

If  $\xi_1=0, \xi_2=0, \xi_3=0$, then the matrix $(h_{l\,m})$ can be chosen to be
diagonal for all $\tau$, and then we say that $g$ has diagonal form.

In this paper we mainly study the non-diagonal case.

To compute the curvature tensor we choose a basis for $\bigwedge^2$ 
\begin{align*}
\{\Omega_1^+,\Omega_2^+,\Omega_3^+,\Omega_1^-\Omega_2^-,\Omega_3^-\},
\end{align*}
where
\begin{align*}
\Omega_1^+&=a^2bc \,dt\wedge\tilde{\sigma}_1+bc \,\tilde{\sigma}_2\wedge\tilde{\sigma}_3 ,\\
\Omega_2^+&=ab^2c \,dt\wedge\tilde{\sigma}_2+ca \,\tilde{\sigma}_3\wedge\tilde{\sigma}_1 ,\\
\Omega_3^+&=abc^2 \,dt\wedge\tilde{\sigma}_3+ab \,\tilde{\sigma}_1\wedge\tilde{\sigma}_2 ,\\
\Omega_1^-&=a^2bc \,dt\wedge\tilde{\sigma}_1-bc \,\tilde{\sigma}_2\wedge\tilde{\sigma}_3 ,\\
\Omega_2^-&=ab^2c \,dt\wedge\tilde{\sigma}_2-ca \,\tilde{\sigma}_3\wedge\tilde{\sigma}_1 ,\\
\Omega_3^-&=abc^2 \,dt\wedge\tilde{\sigma}_3-ab \,\tilde{\sigma}_1\wedge\tilde{\sigma}_2. 
\end{align*}
With respect to this frame, the curvature tensor has the following block form \cite{Besse}
\begin{align*}
\left(
  \begin{array}{rr}
    A   &   B \\
    {}^tB   & D   \\
  \end{array}
\right),
\end{align*}
where $s=4 \,\textrm{trace}D$ is the scalar curvature, 
$W^+=A-\frac{1}{12}\,s$ and $W^-=D-\frac{1}{12}\,s$ are the self-dual and 
anti-self-dual parts of 
the Weyl tensor and $B$ is the trace free parts of Ricci tensor.

We set $w_1=bc, w_2=ca, w_3=ab$ and determine $\alpha_1,\alpha_2, \alpha_3$ by
\begin{align}
\begin{split}
\dot{w}_1&=-w_2 w_3 + w_1(\alpha_2+\alpha_3),\\
\dot{w}_2&=-w_3 w_1 + w_2(\alpha_3+\alpha_1), \\
\dot{w}_3&=-w_1 w_2 + w_3(\alpha_1+\alpha_2).
\end{split}\label{sd1}
\end{align}
Calculating the condition $A=0$, we have the following theorem.

\begin{thm}
The metric is anti-self-dual with vanishing scalar curvature if and only if
$\alpha_1,\alpha_2,\alpha_3$ and $\xi_1,\xi_2,\xi_3$ satisfies the following equations:
\begin{align}
\begin{split}
\dot{\alpha}_1=&-\alpha_2\alpha_3+\alpha_1(\alpha_2+\alpha_3)+
      \frac{1}{4}(w_2^2-w_3^2)^2\left(\frac{\xi_1}{w_2w_3}\right)^2\\
      &{}+
      \frac{1}{4}(w_3^2-w_1^2)(3w_1^2+w_3^2)\left(\frac{\xi_2}{w_3w_1}\right)^2
\\
      &{}+
      \frac{1}{4}(w_2^2-w_1^2)(3w_1^2+w_2^2)\left(\frac{\xi_3}{w_1w_2}\right)^2,
\\
\dot{\alpha}_2=&-\alpha_3\alpha_1+\alpha_2(\alpha_3+\alpha_1)+
      \frac{1}{4}(w_3^2-w_1^2)^2\left(\frac{\xi_2}{w_3 w_1}\right)^2\\
      &{}+
      \frac{1}{4}(w_1^2-w_2^2)(3w_2^2+w_1^2)\left(\frac{\xi_3}{w_1 w_2}\right)^2
\\
      &{}+
      \frac{1}{4}(w_3^2-w_2^2)(3w_2^2+w_3^2)\left(\frac{\xi_1}{w_2w_3}\right)^2,
\\
\dot{\alpha}_3=&-\alpha_1\alpha_2+\alpha_3(\alpha_1+\alpha_2)+
      \frac{1}{4}(w_1^2-w_2^2)^2\left(\frac{\xi_3}{w_1w_2}\right)^2\\
      &{}+
      \frac{1}{4}(w_2^2-w_3^2)(3w_3^2+w_2^2)\left(\frac{\xi_1}{w_2w_3}\right)^2
\\
      &{}+
      \frac{1}{4}(w_1^2-w_3^2)(3w_3^2+w_1^2)\left(\frac{\xi_2}{w_3w_1}\right)^2,
\end{split}\label{sd2}
\end{align}
and
\begin{align}
\begin{split}
(w_2^2-w_3^2)\frac{d}{dt}\left(\frac{\xi_1}{w_2w_3}\right)=&
      \frac{\xi_2}{w_3w_1}\frac{\xi_3}{w_1w_2}(-2w_2^2w_3^2+w_3^2w_1^2+w_1^2w_2^2)\\
      &{}+
      \frac{\xi_1}{w_2w_3}(\alpha_2w_2^2-\alpha_3w_3^2+3\alpha_2w_3^2+3\alpha_3w_2^2),\\
(w_3^2-w_1^2)\frac{d}{dt}\left(\frac{\xi_2}{w_3 w_1}\right)=&
      \frac{\xi_3}{w_1 w_2}\frac{\xi_1}{w_2w_3}(-2w_3^2w_1^2+w_1^2w_2^2+w_2^2w_3^2)\\
      &{}+
      \frac{\xi_2}{w_3w_1}(\alpha_3w_3^2-\alpha_1w_1^2+3\alpha_3w_1^2+3\alpha_1w_2^2),\\
(w_1^2-w_2^2)\frac{d}{dt}\left(\frac{\xi_3}{w_1w_2}\right)=&
      \frac{\xi_1}{w_1w_3}\frac{\xi_2}{w_3w_1}(-2w_1^2w_2^2+w_2^2w_3^2+w_3^2w_1^2)\\
      &{}+
      \frac{\xi_3}{w_1w_2}(\alpha_1w_1^2-\alpha_2w_2^2+3\alpha_1w_2^2+3\alpha_2w_1^2).
\end{split}\label{sd3}
\end{align}
\end{thm}

\begin{rem}
If $\xi_1=0$, $\xi_2=0$ and $\xi_3=0$ then the system \eqref{sd1}, 
\eqref{sd2}, \eqref{sd3} reduce to a sixth-order system
given by Tod\cite{Tod}. Furthermore, if $\alpha_1=w_1, \alpha_2=w_2, \alpha_3=w_3$ 
then \eqref{sd1},\eqref{sd2},\eqref{sd3} reduce to a third-order 
system which determines Atiyha-Hitchin family \cite{AH}, and 
if $\alpha_1=0, \alpha_2=0, \alpha_3=0$ then 
the system reduce to a third-order system which determines
BGPP family \cite{BGPP}.
\end{rem}

\begin{rem}
If $w_2=w_3$, then we can set $\xi_1=0$, $\xi_2=0$ and $\xi_3=0$ by 
taking another flame. This
is also a diagonal case. Therefore we assume $(w_2-w_3)(w_3-w_1)(w_1-w_2)\neq 0$.
\end{rem}


\section{The Isomonodromic Deformations and Painlev\'e equation} \label{Painleve}
Let $(M,g)$ be an oriented Riemannian four manifold. We define $Z$ to be the
unit sphere bundle in the bundle of self-dual two-forms, and let $\pi: Z\to M$
denote the projection. Each point $z$ in the fiber over $\pi(z)$ defines a 
complex structure on the tangent space $T_{\pi(z)}M$, compatible with the metric
and its orientation.

Using the Levi-Civita connection, we can split the tangent space $T_z Z$ into 
horizontal and vertical spaces, and the projection $\pi$ identifies the horizontal
space with $T_{\pi(z)}M$. This space has a complex structure defined by $z$ and 
the vertical space is the tangent space of the fiber $S^3\cong \mathbb{CP}^1$ 
which has its natural 
complex structure. 
The almost complex structure on $Z$ is integrable if and only if the metric is
anti-self-dual \cite{AHS, Penrose}. In this situation $Z$ is called 
the twistor space of $(M,g)$ and 
The fibers are called the real twistor lines.

The almost complex structure 
on $Z$ can be determined by the following
$(1,0)$-forms:
\begin{align}
\begin{split}
\Theta_1=&z(e^2+\sqrt{-1}e^3)-(e^0+\sqrt{-1}e^1),\\
\Theta_2=&z(e^0-\sqrt{-1}e^1)+(e^2-\sqrt{-1}e^3),\\
\Theta_3=&dz + \frac{1}{2}z^2(\omega^0_2-\omega^3_1+\sqrt{-1}(\omega^0_3-\omega^1_2))\\
	&{}-\sqrt{-1}z(\omega^0_1-\omega^2_3)
	+\frac{1}{2}(\omega^0_2-\omega^3_1-\sqrt{-1}(\omega^0_3-\omega^1_2)),
\end{split}
\end{align}
where $\{e^0,e^1,e^2,e^3\}$ is an orthonormal flame, and $\omega^i_j$ are the
connection forms determined by $d e^i+\omega^i_j\wedge e^j=0$ and
$ \omega^i_j+\omega^j_i=0$.
Then the anti-self-dual condition is 
\begin{align}
d\Theta_1&\equiv 0,& d\Theta_2&\equiv 0,& d\Theta_3&\equiv 0 
	&(\textrm{mod}\; \Theta_1, \Theta_2,\Theta_3).
\end{align}

If the metric is $SU(2)$ invariant, we can write
\begin{align}
\left(
  \begin{array}{c}
    \Theta_1   \\
    \Theta_2   \\
    \Theta_3   \\
  \end{array}
\right)=
\left(
  \begin{array}{c}
    0   \\
     0  \\
      1 \\
  \end{array}
\right)dz+
\left(
  \begin{array}{c}
    v_1   \\
     v_2  \\
      v_3 \\
  \end{array}
\right)dt+A\left(
  \begin{array}{c}
    \sigma_1   \\
      \sigma_2 \\
      \sigma_3 \\
  \end{array}
\right),
\end{align}
where $v_1=v_1(z,t)$, $v_2=v_2(z,t)$, $v_3=v_3(z,t)$; $A=\left(a_{i\,j}(z,t)\right)_{i,j=1,2,3}$.

If $\textrm{det}A\equiv 0$, then metric is in the BGPP family \cite{BGPP}.

If $\textrm{det}A\neq 0$, then we can write
\begin{align}
\left(
  \begin{array}{c}
    \sigma_1   \\
    \sigma_2   \\
    \sigma_3   \\
  \end{array}
\right)\equiv -A^{-1}\left(
\left(
  \begin{array}{c}
    0   \\
     0  \\
      1 \\
  \end{array}
\right)dz+
\left(
  \begin{array}{c}
    v_1   \\
    v_2   \\
    v_3   \\
  \end{array}
\right)dt
\right)=:\left(
  \begin{array}{c}
    \varsigma_1   \\
    \varsigma_2   \\
    \varsigma_3   \\
  \end{array}
\right),
\end{align}
and then 
\begin{align}
d\left(
  \begin{array}{c}
    \varsigma_1   \\
    \varsigma_2   \\
    \varsigma_3   \\
  \end{array}
\right)\equiv \left(
  \begin{array}{c}
    \varsigma_2\wedge\varsigma_3   \\
    \varsigma_3\wedge \varsigma_1   \\
    \varsigma_1\wedge \varsigma_2   \\
  \end{array}
\right).
\end{align}
Since $\varsigma_1,\varsigma_2,\varsigma_3$ are one-forms on $(z,t)-$plane,
\begin{align}
d\left(
  \begin{array}{c}
 \varsigma_1      \\
 \varsigma_2      \\
 \varsigma_3      \\
  \end{array}
\right)=\left(
  \begin{array}{c}
    \varsigma_2\wedge \varsigma_3   \\
    \varsigma_3\wedge \varsigma_1   \\
    \varsigma_1\wedge \varsigma_2   \\
  \end{array}
\right).
\end{align}
If we set
\begin{align}
\Sigma&=\frac{1}{\sqrt{2}}\left(
  \begin{array}{cc}
    \sqrt{-1}\varsigma_1   &\varsigma_3+\sqrt{-1}\varsigma_{2}    \\
    \varsigma_3+\sqrt{-1}\varsigma_2   &\sqrt{-1}\varsigma_1    \\
  \end{array}
\right)\\
&=:{}-B_1\,dz-B_2\,dt,
\end{align}
then
\begin{align}
d\Sigma+\Sigma\wedge\Sigma=0.
\end{align}
This is the isomonodromic condition of the equation
\begin{align}
\left(\frac{d}{dz}-B_1\right)\left(
  \begin{array}{c}
    y_1   \\
     y_2  \\
  \end{array}
\right)=0.
\end{align}
$B_1$ has poles on $\{z\,|\,\textrm{det}A=0\}$.
\begin{lemma}\label{detAlemma}
$\textrm{det}\,A=0$ is equivalent to the following equation
\begin{multline}
z^4
\left(
	\left(\alpha_2+\alpha_3
	\right)-\sqrt{-1}X_1
\right)-2z^3
\left(
	X_2-\sqrt{-1}X_3
\right)+2z^2
\left(
	-2\alpha_1+\alpha_2+\alpha_3
\right)\\{}
+2z
\left(
	X_2+\sqrt{-1}X_3
\right)+
\left(
	\left(\alpha_2+\alpha_3
	\right)+\sqrt{-1}X_1
\right)=0,\label{detA}
\end{multline}
where
\begin{align*}
X_1&=\frac{w_2^2-w_3^2}{w_2 w_3}\xi_1,&
X_2&=\frac{w_3^2-w_1^2}{w_3 w_1}\xi_2,&
X_3&=\frac{w_1^2-w_2^2}{w_1 w_2}\xi_3.&
\end{align*}
\end{lemma}
For this lemma, generically $B_1$ has four simple poles.
\begin{thm}
The anti-self-dual equation on $SU(2)-$invariant metrics generically reduce to a 
Painlev\'e VI type equation.
\end{thm}
\begin{rem}
If $z=\zeta$ is a solution of the equation 
then $z=-1/\bar{\zeta}$ is also a solution. Therefore the equation is compatible
with the real structure of twistor space.
\end{rem}
\begin{rem}
The idea of Hitchin $\cite{Hitchin}$ is that the lifted action 
of $SU(2)$ on the twistor space $Z$ gives
a homomorphism of vector 
bundles $\alpha: Z \times \mathfrak{su}(2)^\mathbb{C}\to TZ$, and
the inverse of $\alpha$ gives a flat meromorphic $SL(2,\mathbb{C})$-connection, which
determine isomonodromic deformations. we can think that one-forms 
$\Theta_1,\Theta_2,\Theta_3$ on $Z$ are infinitesimal variations, 
therefore we can identify
$\Sigma$ with $\alpha^{-1}$.
\end{rem}
\begin{lemma}\label{order2lemma}
Let $g$ be a non-diagonal $SU(2)$-invariant metric. Then \eqref{detA} has two solutions of 
order two if and only if there exists a function $f(t)$ satisfying
\begin{align*}
X_1^2=&4(f-\alpha_2)(f-\alpha_3),\\
X_2^2=&4(f-\alpha_3)(f-\alpha_1),\\
X_3^2=&4(f-\alpha_1)(f-\alpha_2).
\end{align*}
And then the anti-self-dual equation reduce to \eqref{sd1}, \eqref{sd2} and $\dot{f}=f^2$.
\end{lemma}

Proof.

We can write \eqref{detA} as 
\begin{align}
\bar{a}\,z^4-\bar{b}\,z^3+c \,z^2+b\,z+a=0,
\end{align}
where $a,b$ are complex coefficient and $c$ is a real coefficient. By an
linear fractional transformation
\begin{align}
z\mapsto \frac{\left(b-|b|\right)\zeta-b+|b|}{\left(-\bar{b}+|b|\right)\zeta-\bar{b}+|b|}
\end{align}
preserving the real structure, we can write \eqref{detA} as 
\begin{align}
\zeta^4-\bar{b}_0\zeta^3+c_0\zeta^2+b_0\zeta+1=0,
\end{align}
where $b_0$ is a complex coefficient and $c_0$ is a real coefficient.
Since this equation is also compatible with the real structure,
if $\zeta=\zeta_0$ is a solution of order two then $\zeta=-1/\bar{\zeta}_0$ is 
also a solution of order two.
Therefore
\begin{align}
\zeta^4-\bar{b}_0\zeta^3+c_0\zeta^2+b_0\zeta+1=(\zeta-\zeta_0)^2(\zeta+1/\bar{\zeta}_0)^2,
\end{align}
then we have $\zeta_0^2\left(-1/\bar{\zeta}_0\right)^2=1$ and 
then $\zeta_0=\pm\bar{\zeta}_0$, which implies $\zeta_0$ is real or 
pure-imaginary. Therefore $b_0$ must be real or pure-imaginary. 
Calculating this condition, we have the Lemma.


\section{The Hermitian Structure} \label{Hermitian}
Hitchin \cite{Hitchin} shows that if a metric is scalar-flat K\"ahler but not Hyper-K\"ahler,
then the anti-self-dual equation reduce to a Painlev\*e III type equation. We can interprets
this result as the following result.
\begin{cor}
If a metric is scalar-flat-K\"ahler but not hyper-K\"ahler then 
the equation \eqref{detA} has two double zeros.
\end{cor}
Therefore we analyze the case \eqref{detA} has two double zeros.

Let $z=z(t)$ is a solution of \eqref{detA}. 
If we restrict $(1,0)$-forms $\Theta_1,\Theta_2$ on $Z$  to $z=z(t)$,
we have $(1,0)$-forms on $M$, which determine
an almost complex structure on $M$. 
Analyzing this almost complex structure, we have the
following theorem.
\begin{thm}\label{hermitian}
Let $g$ be an $SU(2)$-invariant anti-self-dual scalar-flat metric.
There exists a $SU(2)$-invariant hermitian structure $(g,I)$ if and only if
\eqref{detA} has solutions of order two.
\end{thm}

\textit{proof.}

Let $(g,i)$ be a $SU(2)$-invariant hermitian structure.
The complex structure $I$ is determined by $(1,0)$-forms 
$\left.\Theta_1\right|_{z=z(t)}$ and $\left.\Theta_2\right|_{z=z(t)}$, where
$z=z(t)$ is a function on $M$ depending on $t$ only.
Since the complex structure is integrable, $\left.\Theta_3\right|_{z=z(t)}\equiv0$
 $(\textrm{mod}\;\left.\Theta_1\right|_{z=z(t)},\left.\Theta_2\right|_{z=z(t)})$.
Therefore we have
\begin{multline}
\left.dz\right|_{z=z(t)}=\Bigl\{
\frac{1}{4}\left(-\alpha_2+\alpha_3+\sqrt{-1}X_1\right)z^3
\\-\frac{1}{2}\left(
\frac{w_1^2}{w_3^2-w_1^2}X_2+\sqrt{-1}\frac{w_1^2}{w_1^2-w_2^2}X_3
\right)z^2
+\frac{\sqrt{-1}}{2}X_1 z
\\-\frac{1}{2}\left(
\frac{w_1^2}{w_3^2-w_1^2}X_2-\sqrt{-1}\frac{w_1^2}{w_1^2-w_2^2}X_3
\right)\\+
\frac{1}{4}\left(\alpha_2-\alpha_3+\sqrt{-1}X_1\right)z^3
\Bigr\}dt \label{dz/dt}
\end{multline}
On the other hand, since $\left.\Theta_3\right|_{z=z(t)}\equiv0$,
$z=z(t)$ is a solution of \eqref{detA}. Moreover if we substitute
$z=z(t)$ and \eqref{dz/dt} into the derivative of left hand side of \eqref{detA},
it also becomes zero. Therefore \eqref{detA} has solutions of order two.

Conversely, let $z=z_0$ be a solution of order two, then from lemma \ref{detAlemma} we have
\begin{align}
z_0=\frac{X_2X_3\pm\sqrt{X_2^2X_3^2+X_3^2X_2^2+X_1^2X_2^2}}{X_1(X_2-\sqrt{-1}X_3)},\label{z_01}
\end{align}
if $X_1X_2X_3\neq 0$. And then we have $\left.\Theta_3\right|_{z=z(t)}\equiv0$. Therefore
the almost complex structure determined by the $(1,0)$-forms
$\left.\Theta_1\right|_{z=z(t)}$ and $\left.\Theta_2\right|_{z=z(t)}$ is
integrable. 

If $X_1X_2X_3=0$, $f$ must be $\alpha_1,\alpha_2$ or $\alpha_3$. 
Let $f=\alpha_1$, then we have $X_2 = 0$ and $X_3=0$, and then 
\begin{align}
z_0=\frac{\sqrt{\alpha_3-\alpha_1}+
\sqrt{-1}\sqrt{\alpha_2-\alpha_1}}{\sqrt{\alpha_2+\alpha_3+2\alpha_1}},\label{z_02}
\end{align}
and then $\left.\Theta_3\right|_{z=z(t)}\equiv0$. In this 
case the almost complex structure is also integrable.


\begin{thm}
The hermitian structure $(g,I)$ determined by theorem $\ref{hermitian}$ is K\"ahler 
if and only if 
\begin{align}
X_1^2&=4\alpha_2\alpha_3,& 
X_2^2&=4\alpha_3\alpha_1,&
X_3^2&=4\alpha_1\alpha_2.
\end{align}
\end{thm}

\textit{proof}.

If $X_1X_2X_3\neq 0$, the K\"ahler form is determined by \eqref{z_01} as
\begin{align*}
\Omega=&\frac{X_2X_3}{\sqrt{X_2^2X_3^2+X_3^2X_2^2+X_1^2X_2^2}}\Omega_1^+\\
&{}+\frac{X_3X_1}{\sqrt{X_2^2X_3^2+X_3^2X_2^2+X_1^2X_2^2}}\Omega_2^+\\
&{}+\frac{X_1X_2}{\sqrt{X_2^2X_3^2+X_3^2X_2^2+X_1^2X_2^2}}\Omega_3^+.
\end{align*}
By the anti-self-dual equations \eqref{sd1},\eqref{sd2},\eqref{sd3}, we have
\begin{align*}
d\Omega=&\frac{2f\,w_1X_2X_3}{\sqrt{X_2^2X_3^2+X_3^2X_2^2+X_1^2X_2^2}}\,
dt\wedge\tilde{\sigma}_2\wedge\tilde{\sigma}_3\\
&{}+\frac{2f\,w_2X_3X_1}{\sqrt{X_2^2X_3^2+X_3^2X_2^2+X_1^2X_2^2}}\,
dt\wedge\tilde{\sigma}_3\wedge\tilde{\sigma}_1\\
&{}+\frac{2f\,w_3X_1X_2}{\sqrt{X_2^2X_3^2+X_3^2X_2^2+X_1^2X_2^2}}\,
dt\wedge\tilde{\sigma}_1\wedge\tilde{\sigma}_2.
\end{align*}
Since $w_1w_2w_3\neq 0$ and $X_1X_2X_3\neq 0$, we have $d\Omega=0$ if and only if $f=0$.

If $X_1X_2X_3=0$, then $f$ must be $\alpha_1,\alpha_2$ or $\alpha_3$. Let
$f=\alpha_1$, then $X_1^2=4(\alpha_2-\alpha_1)(\alpha_3-\alpha_1)$, $X_2=0$, $X_3=0$.
The K\"ahler form is determined by \eqref{z_02} as
\begin{align}
\Omega=\frac{\sqrt{\alpha_2-\alpha_1}}{\sqrt{\alpha_2+\alpha_3-2\alpha_1}}\Omega_2^+
+\frac{\sqrt{\alpha_3-\alpha_1}}{\sqrt{\alpha_2+\alpha_3-2\alpha_1}}\Omega_3^+.
\end{align}
Then 
\begin{align}
d\Omega=\frac{2w_2\alpha_1\sqrt{\alpha_2-\alpha_1}}{\sqrt{\alpha_2+\alpha_3-2\alpha_1}}
dt\wedge\tilde{\alpha}_3\wedge\tilde{\alpha}_1+
\frac{2w_3\alpha_1\sqrt{\alpha_3-\alpha_1}}{\sqrt{\alpha_2+
\alpha_3-2\alpha_1}}dt\wedge\tilde{\alpha}_1\wedge{\alpha}_2.
\end{align}
Since the metric is non-diagonal, $X_1^2=4(\alpha_2-\alpha_1)(\alpha_3-\alpha_1)\neq 0$
and then $d\Omega=0$ if and only if $\alpha_1=0$.
\begin{rem}
If the metric is scalar-flat K\"ahler, the anti-self-dual equation reduce to a sixth-order
equation.
\end{rem}

\end{document}